\documentclass[prx,aps,twocolumn,longbibliography]{revtex4-1}
\usepackage[utf8]{inputenc}
\usepackage{xcolor}
\usepackage{amsmath,amsthm}
\usepackage{amsfonts}
\usepackage{amssymb}
\usepackage{makeidx}
\usepackage{graphicx}
\usepackage{tikz}
\usepackage{tikz-cd}
\usepackage{braket}
\usepackage[caption=false]{subfig}
\usepackage{color}
\usepackage[colorlinks=true,citecolor=black,linkcolor=black,urlcolor=black]{hyperref}
\usepackage{url}
\DeclareMathOperator{\tr}{\mathrm{tr}}

\newcommand{\kb}{k_\mathrm{B}}

\theoremstyle{remark}

\begin{document}

\title{The role of entropy production and thermodynamic uncertainty relations in the asymmetric thermalization of open quantum systems}
\author{\'Alvaro Tejero}\email[]{atejero@onsager.ugr.es}\affiliation{Quantum Thermodynamics and Computation Group, Departamento de Electromagnetismo y Física de la Materia, Universidad de Granada, 18071 Granada, Spain\looseness=-1}
\affiliation{Instituto Carlos I de Física Teórica y Computacional, Universidad de Granada, 18071 Granada, Spain\looseness=-1}

\begin{abstract}
The asymmetry between heating and cooling in open quantum systems is a hallmark of nonequilibrium dynamics. Here, we investigate the thermalization of a quantum system weakly coupled to a bosonic thermal bath under detailed balance, focusing on the entropy production rate and the quantum thermokinetic uncertainty relation (TKUR). Within the sector of states that remain diagonal in the energy basis we derive a perturbative expression for the time-dependent entropy production rate by expanding the relative entropy around equilibrium in terms of Liouvillian eigenmodes. 
Through this expansion, we determine that the spectral gap, the overlap of the initial state with the slowest decay mode, and the mode weight control the entropy production. 
Particularly for the thermal qubit, the ratio of the initial rates of heating and cooling has an exact closed form. 
Heating involves a higher initial entropy production, and hence faster thermalization, solely because of its larger spectral gap, i.e., the bath-temperature dependence of the dissipative rate.
The TKUR is then used as a diagnostic tool that relates the asymmetry to the precision bounds on heat-current fluctuations. 
The analysis presented here is intended to clarify the thermodynamic origin of the asymmetry within the well-controlled Markovian case.
\end{abstract}

\maketitle

\section{Introduction}
The dynamics of open quantum systems far from thermal equilibrium exhibits complex and often counterintuitive behavior \cite{binder2018,strasberg2022,divita2022}, particularly in transient regimes where standard linear response theory \cite{kubo1957,kubo1966,konopik2019,blair2024} and fluctuation theorems \cite{jarzynski1997,crooks1999,campisi2021} fail to fully capture the evolution. 
Understanding the general behavior of systems out of thermal equilibrium is essential both for comprehending the underlying physics, and for enabling applications in areas such as quantum information processing \cite{goold2016,strasberg2017,deffner2019} or quantum thermal machines \cite{bhattacharjee2021,tejero2024,tejero2024squeezing,dunlop2025}.

Striking manifestations of nonequilibrium behavior in quantum systems, such as the quantum Mpemba effect \cite{lu2017, carollo2021,nava2024, strachan2024, zhang2025, vanvu2025_mpemba} and the asymmetry between heating and cooling processes in open quantum systems \cite{lapolla2020,vanvu2021,ibanez2024,tejero2025}, where heating typically occurs more rapidly than cooling, are closely tied to the thermalization dynamics. 
The so-called \emph{thermal kinematics} framework \cite{ibanez2024,tejero2025}, which integrates quantum information geometry with thermodynamics, provides an effective characterization of the temporal evolution of these phenomena across diverse scenarios \cite{crooks2007,deffner2017,bravetti2025,tejero2025,bettmann2025}.
Ref. \cite{tejero2025} provides a concise mathematical explanation of the asymmetry between heating and cooling processes in open quantum systems from both mathematical and dynamical perspectives. In that work, the spectra of the corresponding Liouvillian operators are compared, with particular attention to the decaying modes associated with real eigenvalues. 
The larger spectral gap of the Liouvillian---defined as the distance to its first nonzero real eigenvalue---observed in the heating case, combined with the influence of the initial state in the dynamics, provides a clear rationale for the phenomena reported.

Nevertheless, a fully thermodynamic account of the asymmetry, expressed in terms of entropy production and its relation to current precision, has not yet been developed systematically.
Although the combination of information geometry and thermodynamics offers an operational explanation for why heating is faster than cooling, it still lacks a direct link to standard nonequilibrium thermodynamic quantities. 
This gap can be bridged by extending the thermal kinematics framework to include the entropy production rate \cite{landi2021,strasberg2022,seifert2025}, a standard quantity in nonequilibrium classical and quantum thermodynamics, to clarify the underlying mechanisms. This approach mirrors that used for classical systems \cite{vanvu2021}. 
Furthermore, we incorporate the quantum uncertainty relations \cite{hasegawa2021} into the analysis, in particular the quantum thermokinetic uncertainty relation (TKUR) \cite{vanvu2025}, which unifies the quantum thermodynamic uncertainty relation (TUR) \cite{barato2015,pietzonka2016,horowitz2020} and the quantum kinetic uncertainty relation (KUR) \cite{garrahan2017,diterlizzi2019} into a single bound. 
With this in mind, the present work provides (i) a spectral-mode expansion of the entropy production rate highlighting the contribution of the spectral gap and the initial-state overlap, (ii) an exact closed-form expression for the ratio of heating to cooling rates in the thermal qubit, which shows that the asymmetry is carried entirely by the spectral gap, and (iii) a diagnostic use of the TKUR that relates the observed asymmetry to bounds on heat-current precision, while distinguishing bounds from actual fluctuations.

To compare these processes, a quantum system weakly coupled to a single bosonic bath is analyzed under the detailed-balance condition, restricting attention to the commuting sector of diagonal states under Davies dynamics. The spectral decomposition of the Liouvillian \cite{albert2014,cavina2017,minganti2018,minganti2019} is examined and related to the entropy production rate, a quantity directly connected to the speed of evolution toward the equilibrium stationary state. The analysis identifies the spectral gap, the overlap with the slowest decay mode, and a mode weight as the fundamental quantities governing the entropy production rate.
These components are evaluated exactly in an analytically tractable model, the thermal qubit, for which the ratio of the initial rates admits a compact closed form.
This result confirms that the entropy production rate is larger for heating than for cooling, in line with the faster thermalization in the heating case, and we find that the asymmetry originates from the larger spectral gap alone, in this case.
Building on this result, a subsequent dedicated section applies the quantum TKUR as a tool that clarifies how the entropy production rate relates to the precision bounds on the dynamics in each case. 
This analysis provides a thermodynamic perspective on the observed processes, emphasizing the origin of the asymmetry through the interplay between entropy production and the quantum TKUR.

The paper is structured as follows. The first section introduces the general framework and dynamics of open quantum systems weakly coupled to a bosonic thermal bath, followed by a discussion of the entropy production rate and the uncertainty relations. 
General calculations of the entropy production rates for heating and cooling processes under the conditions described above, along with their comparison, are then presented. 
This is followed by the application to a thermal qubit, offering a physical explanation and justification for the observed asymmetry.
A subsequent section on quantum uncertainty relations further clarifies the asymmetry by relating entropy production and dynamical activity to fundamental quantum limits. The paper concludes with a summary of the main findings.

\section{Preliminaries}
\subsection{Dynamics}\label{sec:dynamics}
We aim to study the temporal evolution of a quantum system evolving from a thermal state at an initial inverse temperature $\beta_0=1/(k_\mathrm{B}T_0)$ to a thermal state at a final inverse temperature $\beta=1/(k_\mathrm{B}T)$. To extract thermodynamic insights, we adopt the simplest and most direct approach.
Specifically, the system is weakly coupled to the environment, and its dynamics is governed by the Gorini–Kossakowski–Sudarshan–Lindblad (GKSL) equation \cite{gorini1976,lindblad1976,manzano2020}. 
In the heating case, the system evolves from an initial inverse temperature $\beta_0 = \beta_\text{C}$ to $\beta = \beta_\text{H} < \beta_\text{C}$. Conversely, in the cooling case, the evolution proceeds from $\beta_0 = \beta_\text{H}$ to $\beta = \beta_\text{C}$.

The state of the system, $\rho$, evolves according to the GKSL master equation
\begin{equation}\label{eqn:lindblad}
\frac{d\rho}{dt}= -\frac{i}{\hbar} [H, \rho] + \sum_{k=1}^N \left( L_k \rho L_k^\dagger - \frac{1}{2} \{ L_k^\dagger L_k, \rho \} \right) \equiv \mathcal{L}[\rho],
\end{equation}
where $H$ is the system Hamiltonian, and the jump operators $L_k$, for $k=1,\ldots,N$, model the energy exchange with the environment. 
The entire right-hand side of Eq.~(\ref{eqn:lindblad}) is the Lindbladian superoperator, $\mathcal{L}$, acting on the state of the system, $\rho$. 
With a single bosonic thermal bath at $\beta$, the Lindbladian dynamics drives the system toward the corresponding Gibbs state $\tau_{\beta}$, meaning that the system evolves asymptotically to thermal equilibrium
\begin{equation}\label{eqn:gibbs}
    \rho(t \to \infty) \equiv \tau_{\beta} = \dfrac{1}{Z}e^{-\beta H},
\end{equation}
where $Z = \tr[\exp(-\beta H)]$ is the partition function, and the steady-state condition $ \mathcal{L}[\tau_{\beta}] = 0 $ holds. 

The Lindbladian acts directly on density operators, as defined in Eq.~(\ref{eqn:lindblad}). Assuming it is diagonalizable, there exists a set of right eigenmatrices $\{\Lambda_j^r\}_{j=0}^{d^2-1}$ \cite{manzano2018}, where $d$ is the dimension of the Hilbert space, satisfying
\begin{equation}
\mathcal{L}\left[\Lambda_j^{r}\right] = \lambda_j \Lambda_j^{r},
\end{equation}
with $\lambda_j$ being the corresponding eigenvalues, for $j=0,\ldots, d^2-1$.
Each right eigenmatrix has an associated left eigenmatrix, $\Lambda_j^{\ell}$, which diagonalizes the conjugated Lindbladian $\mathcal{L}^\dagger$, i.e.,
\begin{equation}
    \mathcal{L}^\dagger\left[\Lambda_j^{\ell}\right] = \lambda_j \Lambda_j^{\ell}.
\end{equation}
The set of left and right eigenmatrices satisfies the orthogonality condition $\tr(\Lambda_i^{\ell}\Lambda_j^{r}) = \delta_{ij}$. 
As the dynamics is completely positive and trace-preserving, the eigenvalues have $\text{Re}(\lambda_j) \leq 0$, for all $j=0,\ldots,d^2 - 1$, ensuring decay of nonequilibrium modes. These eigenvalues are classified into three different groups \cite{manzano2018}: (i) nonzero real eigenvalues, representing the \emph{decay modes}; (ii) pair of complex-conjugate states, representing the \emph{oscillating coherences}; and (iii), the null eigenvalue $\lambda_0 = 0$, which corresponds to $\Lambda_0^r = \tau_{\beta}$, i.e., the \emph{stationary state}. 
The existence of a null eigenvalue in the Lindbladian dynamics is ensured by Evans theorem \cite{evans1997,albert2014}.
For all $j\neq 0$, the right eigenmatrices are traceless, $\tr(\Lambda_j^r) = 0$ \cite{minganti2018,minganti2019}.

For any initial state $\rho_0$, its time evolution can be expressed in terms of the Lindbladian spectral decomposition as
\begin{equation}\label{eqn:eigenexpansion}
    \rho(t) = \tau_{\beta} + \sum_{j=1}^{d^2 -1} c_j e^{\lambda_j t} \Lambda_j^{r},
\end{equation}
where the coefficient $c_j = \tr(\Lambda_j^\ell \rho_0)$ quantifies the overlap of the initial state with the $j$th mode, for $j=1,\ldots,d^2-1$. 
Given the spectral decomposition, the spectral gap is defined as the absolute value of the real part of the nonzero eigenvalue closest to the origin,
\begin{equation}\label{eqn:spectral_gap}
\left|\operatorname{Re}(\lambda_1)\right| = \min_{j\neq 0}\bigl|\operatorname{Re}(\lambda_j)\bigr|,
\end{equation}
where $\lambda_1$ labels the slowest decaying mode. 
This gap sets the asymptotic thermalization timescale $\tilde{t}\sim 1/\bigl|\operatorname{Re}(\lambda_1)\bigr|$.
It is worth noting that more complex settings, beyond thermalization with a single weakly coupled thermal bath, may follow a different form of this relation. 
Nevertheless, in all cases, the relaxation time remains inversely proportional to the spectral gap \cite{mori2020,haga2021}.
The spectral gap reflects the bath’s dissipative \emph{strength}, while the overlaps quantify how the initial state couples to each mode. 
In particular, $c_1$ measures the projection of the initial state onto the slowest decaying mode, thereby influencing the effective thermalization speed.

The Lindblad operators in Eq.~(\ref{eqn:lindblad}) are derived under the standard weak-coupling, Born–Markov and secular approximations for a system interacting with a bosonic thermal bath whose spectral density is taken to be flat (or evaluated at the relevant Bohr frequencies and absorbed into a constant rate). Under these assumptions the jump rates are proportional to the Bose–Einstein occupation
\begin{equation}
    n_\mathrm{B}(\beta) = \left[\exp(\beta \hbar \omega) - 1\right]^{-1},
\end{equation}
where $\omega$ denotes the Bohr frequency associated with a given jump operator $A$. The coupling strength $\gamma$ is assumed independent of temperature and of frequency.
Note that this is a common modeling choice but not universal.
Frequency-dependent spectral densities would introduce additional structure for multi-level systems. 
Considering two temperatures, $T_\mathrm{C}$ and $T_\mathrm{H}>T_\mathrm{C}$, the corresponding inverse temperatures satisfy $\beta_\mathrm{H}<\beta_\mathrm{C}$.

Particularly, for the thermal qubit treated in Sec.~\ref{sec:thermal_qubit}, the spectral gap can be computed exactly as $|\text{Re}(\lambda_1)|=\gamma[1+2n_\mathrm{B}(\beta)]$, which is strictly increasing with the bath temperature, since $n_\mathrm{B}(\beta_\mathrm{H})>n_\mathrm{B}(\beta_\mathrm{C})$. 
For a more general case, however, the increase of the individual transition rates $\gamma n_\mathrm{B}(\beta)$ and $\gamma[1+n_\mathrm{B}(\beta)]$ with temperature does not by itself fix the ordering of the gap, since the latter depends on the whole set of transitions and on the stationary state. We therefore formulate the ordering as an explicit assumption, hereafter called (A),
\begin{equation}\label{eqn:assumption_gap}
    \left|\text{Re}\left[\lambda_1(\beta_\mathrm{H})\right]\right| > \left|\text{Re}\left[\lambda_1(\beta_\mathrm{C})\right]\right|
\end{equation}
for $\beta_\mathrm{H}<\beta_\mathrm{C}$. This relation holds exactly for the thermal qubit, and is supported numerically for the models studied in Ref.~\cite{tejero2025}, but it must be verified for each specific model.

In the following, the Lindblad operators take the form
\begin{equation}\label{eqn:lindblad_operators}
    L_1 = \sqrt{\gamma [1 + n_\mathrm{B}(\beta)]} A, \ L_2 = \sqrt{\gamma n_\mathrm{B}(\beta)} A^\dagger,
\end{equation}
where $A$ is the jump operator associated with the Bohr frequency $\omega$. 
The dissipative rates scale as $\gamma(1+n_\mathrm{B})$ for de-excitation and $\gamma n_\mathrm{B}$ for excitation, and they satisfy the detailed-balance condition
\begin{equation}\label{eqn:detailed_balance}
    \frac{\gamma n_\text{B}(\beta)}{\gamma [1 + n_\text{B}(\beta)]} = e^{-\beta \hbar \omega},
\end{equation}
which ensures thermalization to $\tau_{\beta}$.

\subsection{Entropy production rate and thermodynamic uncertainty relations}
For the process described in Sec.~\ref{sec:dynamics}, the second law of thermodynamics can be expressed in terms of the entropy production rate \cite{landi2021}.
Since the system exchanges energy with a thermal bath, this quantity is defined \emph{à la Clausius} as
\begin{equation}\label{eqn:production}
    \dot{\Sigma} \equiv \sigma = \dfrac{d S(\rho)}{dt} - \kb\beta\dfrac{d Q}{dt} = \dfrac{d S(\rho)}{dt} - \dfrac{1}{T}\dfrac{d Q}{dt} \geq 0.
\end{equation}
where $S(\rho) = -\kb\tr(\rho \ln \rho)$ denotes the von Neumann entropy of the system, and the heat flow from the bath is given by $\dot{Q} = \tr(H \mathcal{L}[\rho])$. 
The time derivative of the entropy is related to the Lindbladian in Eq.~(\ref{eqn:lindblad}) via $\dot{S}(\rho) = -\kb\tr(\mathcal{L}[\rho] \ln \rho)$. Equation (\ref{eqn:production}) represents a specific form of the entropy production rate. 
Using $\ln\tau_{\beta}=-\beta H-\ln Z$ and $\tr\mathcal{L}[\rho]=0$, one finds
\begin{equation}
    -\kb\frac{d}{dt}D(\rho\Vert\tau_\beta)=\dfrac{d S(\rho)}{dt}-\kb\beta\dfrac{d Q}{dt},
\end{equation}
so that Eq.~(\ref{eqn:production}) coincides with the general definition given next, given by Eq.~(\ref{eqn:production2}).
Its general definition, for a system evolving toward a fixed thermal state $\tau_{\beta}$, i.e., when the bath temperature is held constant, is
\begin{equation}\label{eqn:production2}
    \sigma := -\kb\frac{d}{dt} D\left(\rho || \tau_{\beta}\right),
\end{equation}
where $D(\rho \Vert \tau_{\beta}) = \tr[\rho (\ln \rho - \ln \tau_{\beta})]$ denotes the relative entropy between the states $\rho$ and $\tau_{\beta}$.

Physically, the relative entropy quantifies how distinguishable $\rho$  is from the equilibrium state $\tau_{\beta}$. 
As the system evolves, $\rho$ approaches $\tau_{\beta}$, so $D(\rho || \tau_{\beta})$ decreases monotonically to zero under the Lindbladian dynamics. 
Its time derivative, $\frac{d}{dt} D(\rho || \tau_{\beta})$, measures the rate at which the state of the system becomes less distinguishable from $\tau_{\beta}$. Since $ D(\rho || \tau_{\beta}) $ decreases, this derivative is negative, making the entropy production rate $\sigma$ positive and representing the \emph{speed} of convergence to equilibrium. Consequently, a higher entropy production rate corresponds to a faster convergence toward the equilibrium state.

In such nonequilibrium processes, fluctuations play a central role. This has motivated the development of the \emph{thermodynamic uncertainty relation} (TUR) \cite{barato2015,pietzonka2016,horowitz2020,strasberg2022}, which connects the precision of reaching a given value to an inherent thermodynamic cost.
The TUR establishes a lower bound on the fluctuations of current-type observables in terms of the entropy production rate. Let $\phi$ denote a time-integrated thermodynamic current. In our case, it represents the heat exchanged with the bath. The uncertainty relation reads
\begin{equation}\label{eqn:tur}
    \tau \dfrac{\mathrm{Var}(\phi)}{\langle \phi\rangle^2} \geq \dfrac{2}{\sigma},
\end{equation}
where $\tau$ is the operational time over which the current is accumulated. The left-hand side is the relative fluctuation $F_\phi$. A larger entropy production rate $\sigma$ lowers the bound, thereby opening the possibility of smaller relative fluctuations, but does not by itself guarantee that the actual fluctuations are reduced. 
The same relative fluctuation can be written
\begin{equation}\label{eqn:tur2}
   F_\phi := \tau\dfrac{\mathrm{Var}(\phi)}{\langle \phi\rangle^2}.
\end{equation}
Similarly, the \emph{kinetic uncertainty relation} (KUR) \cite{garrahan2017,diterlizzi2019} sets a lower bound on the precision of observables associated with dynamical activity in Markovian systems
\begin{equation}\label{kur}
   F_\phi \geq 1/\alpha,
\end{equation}
where $\alpha$ is the \emph{dynamical activity}. This quantity, relevant in both classical and quantum stochastic systems, represents the total number of configuration changes along a trajectory \cite{garrahan2017}. Its extension to the quantum case and to our problem of interest is discussed in Sec. \ref{subsec:uncertainty}. 
The motivation for expressing the TUR as in Eq.~(\ref{eqn:tur2}) is that both bounds can be combined into a single expression, the \emph{quantum thermokinetic uncertainty relation} (TKUR) \cite{vanvu2025}. This bound takes the form
\begin{equation}\label{eqn:tkur}
\frac{F_\phi}{(1 + \delta_\phi)^2} \geq \frac{4\alpha}{\sigma^2} \Phi\left( \frac{\sigma}{2\alpha} \right)^2 \geq \max\left( \frac{2}{\sigma}, \frac{1}{\alpha} \right),
\end{equation}
where $\Phi(x)$ is the inverse function of $x \tanh(x)$, and $\delta_\phi$ is a quantum correction term, which vanishes in the classical limit.

As discussed in Sec.~\ref{sec:dynamics}, we focus on the dissipative dynamics of a system weakly coupled to a single thermal bath, with jump operators specified in Eq.~(\ref{eqn:lindblad_operators}). As noted in \cite{vanvu2025}, Lindbladian dynamics can be regarded as a special case of quantum stochastic dynamics. Consequently, the uncertainty relations in Eqs.~(\ref{eqn:tur})--(\ref{eqn:tkur}) apply to this scenario.
The implications of these relations for thermalization are analyzed later for the analytically solvable case of the thermal qubit.

\section{Quantification of the thermalization speed}
Building on the previous discussion, one may seek a general expression for the time evolution of the entropy production rate, i.e., a way to quantify Eq.~(\ref{eqn:production2}) for a quantum system weakly coupled to a thermal bath. The goal is not to derive a closed-form solution, but to obtain a form suitable for comparing distinct processes, such as heating and cooling. To this end, several approximations are introduced to highlight the underlying physical insights.

\subsection{Approximated expression for the entropy production rate}\label{subsec:approximated_entropyprod}

To compute the entropy production rate analytically, we consider deviations from the final equilibrium state $\tau_{\beta}$. 
We work within the class of Davies-type thermalization dynamics starting from thermal (hence diagonal) initial states. 
Under these conditions the evolved state $\rho(t)$ remains diagonal in the energy eigenbasis and therefore commutes with the thermal state $\tau_{\beta}$ at all times. This means that the logarithm of their product splits into a sum of logarithms without additional correction terms, see Eq.~(\ref{eqn:expansion_delta3}) below. The subsequent analysis is therefore restricted to this commuting sector. Its extension to genuinely non-commuting dynamics, e.g. coherent driving or non-diagonal initial states, would require a more general treatment.

With this restriction in mind, the time evolution of the system state, given by Eq.~(\ref{eqn:eigenexpansion}), can be expressed as
\begin{equation}\label{eqn:expansion_delta}
    \rho(t) = \tau_{\beta} + \Delta \rho(t), 
\end{equation}
where 
\begin{equation}
    \Delta \rho(t) = \sum_{j=1}^{d^2 -1} c_j e^{\lambda_j t} \Lambda_j^{r}
\end{equation}
represents the deviation from the equilibrium state. 
Since $\tau_{\beta}$ is the full rank Gibbs state at inverse temperature $\beta$, Eq.~(\ref{eqn:expansion_delta}) can be written as
\begin{equation}\label{eqn:expansion_delta2}
    \rho(t) = \tau_{\beta}\left[\mathrm{Id} + \tau_{\beta}^{-1} \Delta \rho(t)\right], 
\end{equation}
where $\text{Id}$ is the identity operator. This implies
\begin{equation}\label{eqn:expansion_delta3}
    \ln\rho(t) = \ln \tau_{\beta} + \ln\left[\mathrm{Id} + \tau_{\beta}^{-1} \Delta \rho(t)\right].
\end{equation}
Here the splitting of the logarithm is exact since $\tau_{\beta}$ and $\Delta\rho(t)$ are simultaneously diagonal in the energy basis, so that all the operators involved commute, as noted before. The expansion of the logarithm is controlled by the operator norm of $X\equiv\tau_{\beta}^{-1}\Delta\rho(t)$. Two complementary interpretations are possible. First, because $\Delta\rho(t)$ decays exponentially under the Lindbladian, $\|X\|$ becomes arbitrarily small at sufficiently late times; the expansion is then an asymptotic long-time approximation. Second, for temperature quenches that are not too large, $\|X\|$ remains moderate already at $t=0$, so the same expansion may be used as a small-quench approximation. In either case the second-order truncation
\begin{equation}
\ln(\mathrm{Id}+X)=X-X^{2}/2+\mathcal{O}(X^{3})
\end{equation}
is justified provided $\|X\|\ll1$. At low temperatures the eigenvalues of $\tau_{\beta}^{-1}$ can become exponentially large, which restricts the admissible size of the quench; this limitation should be kept in mind when applying the formula far from the high-temperature regime. Thus,
\begin{equation}
    \ln \rho(t) \approx  \ln \tau_{\beta} + \tau_{\beta}^{-1} \Delta \rho(t) - \frac{1}{2}\left[\tau_{\beta}^{-1} \Delta \rho(t) \right]^2.
\end{equation}
This expansion allows the relative entropy to be approximated by
\begin{equation}
    \begin{split}
        &D\left(\rho || \tau_{\beta}\right)   \approx\\
        &\tr\left\{ \left[\tau_{\beta} + \Delta \rho(t)\right] \left(\tau_{\beta}^{-1} \Delta \rho(t) - \frac{1}{2}\left[\tau_{\beta}^{-1} \Delta \rho(t) \right]^2 \right) \right\}.
    \end{split}
\end{equation}
Using the property $\tr(\Lambda_j^r) = 0$ for all $j \neq 0$, and discarding higher-order terms in the expansion, the only contributions to the relative entropy that remain are
\begin{equation}\label{eqn:relative_eigenmatrices}
\begin{split}
        D\left(\rho || \tau_{\beta}\right) & = \tr\left[\Delta \rho(t) \tau_{\beta}^{-1} \Delta \rho(t)\right] \\
        &- \frac{1}{2} \tr\left(\tau_{\beta} \left[ \tau_{\beta}^{-1} \Delta \rho(t)\right]^2\right) \\
        & = \dfrac{1}{2} \tr\left[ \Delta \rho(t) \tau_{\beta}^{-1} \Delta \rho(t) \right]   \\
        & =  \frac{1}{2} \sum_{j=1}^{m} |c_j|^2 e^{2 \operatorname{Re}(\lambda_j) t} \operatorname{tr}(\Lambda^r_j \tau_{\beta}^{-1} \Lambda^r_j).
        \end{split}
\end{equation}
Here, the sum accounts only for the decay modes, so it runs from $j=1$ to $j=m$, where $m$ is the total number of decaying modes of the Lindbladian ($m=d-1$ in the sector of populations for a nondegenerate spectrum). The last line of Eq.~(\ref{eqn:relative_eigenmatrices}) relies on a weighted orthogonality of the eigenmodes, which does not follow from $\tr(\Lambda_j^r)=0$ alone, a property only used to cancel the term linear in $\Delta\rho$, but from detailed balance as we now show.

Consider the commuting sector. The Lindbladian $\mathcal{L}$ acts on the populations as a rate matrix $L_{ij}$ inducing the transition $j\to i$, such that $L_{ij}\pi_j=L_{ji}\pi_i$, where $\pi_i$ are the eigenvalues of $\tau_\beta$, i.e., Eq.~(\ref{eqn:detailed_balance}) at each Bohr frequency.
Then $\tau_{\beta}^{-1/2}\mathcal{L}\,\tau_{\beta}^{1/2}$ is real symmetric, so $\mathcal{L}$ is self-adjoint with respect to the inner product $\langle A,B\rangle_{\tau}=\tr(A^\dagger\tau_{\beta}^{-1}B)$. The eigenvalues are therefore real and the right eigenmatrices can be chosen orthogonal for this inner product, even within degenerate eigenspaces:
\begin{equation}\label{eqn:weighted_orthogonality}
    \tr\!\left(\Lambda_i^r\tau_{\beta}^{-1}\Lambda_j^r\right)=2W_j \delta_{ij},
\end{equation}
so that the cross terms with $i\neq j$ vanish exactly. 
The left eigenmatrices are $\Lambda_j^\ell=\tau_{\beta}^{-1}\Lambda_j^r/(2W_j)$, which satisfy $\tr(\Lambda_i^\ell\Lambda_j^r)=\delta_{ij}$, and the overlaps $c_j$ are real. 

Note that the product $|c_j|^2W_j$ is invariant under the rescaling $\Lambda_j^r\to a\Lambda_j^r$ (with $c_j\to c_j/a$ and $W_j\to a^2W_j$), whereas $W_j$ alone is not. Comparing weights between different processes therefore requires a common normalization of the eigenmatrices.
For the particualr example of the thermal qubit, see Sec.~\ref{sec:thermal_qubit}, it is fixed by $\Lambda_1^r=(1\ \ {-1})^T$. The modal representation is valid for generators obeying detailed balance and for initial states diagonal in the energy basis. Outside this sector, the weighted orthogonality is not guaranteed.

The relevant term in the sum is the trace $\tr ( \Lambda_j^r \tau_{\beta}^{-1} \Lambda_j^r )$, which involves the eigenmatrices weighted by the inverse of the final thermal state $\tau_{\beta}$.
Equation (\ref{eqn:relative_eigenmatrices}) can be simplified to 
\begin{equation}\label{eqn:relative_eigenmatrices2}
\begin{split}
    D(\rho(t) || \tau_{\beta}) & = \frac{1}{2} \sum_{j=1}^{m} |c_j|^2 e^{2 \operatorname{Re}(\lambda_j) t} \operatorname{tr}(\Lambda^r_j \tau_{\beta}^{-1} \Lambda^r_j)\\
        & = \sum_{j=1}^{m} |c_j|^2 e^{2 \operatorname{Re}(\lambda_j) t} W_j,
\end{split}
\end{equation}
by defining the scalar weight $ W_j :=\tr(\Lambda^r_j \tau_{\beta}^{-1} \Lambda^r_j)/2 $, which quantifies the contribution of the $j$th mode to the quadratic approximation.
The entropy production rate is simply obtained by differentiating Eq.~(\ref{eqn:relative_eigenmatrices2})
\begin{equation}
    \sigma(t) = -\kb\sum_{j=1}^{m} |c_j|^2 [2 \text{Re}(\lambda_j)] e^{2 \text{Re}(\lambda_j) t} W_j,
\end{equation}
which, for the real, negative eigenvalues corresponding to the decaying modes, $\text{Re}(\lambda_j) < 0$, for $j =1, \ldots,m$, leading to
\begin{equation}\label{eqn:dot_sigma}
    \sigma(t) = 2\kb\sum_{j=1}^{m} |c_j|^2 |\text{Re}(\lambda_j)| e^{-2 |\text{Re}(\lambda_j)| t} W_j.
\end{equation}
Focusing on the slowest decaying mode, $j=1$, the dominant contribution is directly given by
\begin{equation}\label{eqn:entropy_production_slowest}
    \sigma(t) =  2\kb |c_1|^2 |\text{Re}(\lambda_1)| e^{-2 |\text{Re}(\lambda_1)| t} W_1.
\end{equation}
In Eq.~(\ref{eqn:dot_sigma}), all terms in the sum are positive, guaranteeing a positive entropy production rate that vanishes as $t \to \infty$, as expected for a system relaxing to the equilibrium state $\tau_{\beta}$. 
The derivation applies to Markovian thermalization processes that preserve diagonality in the energy basis and satisfy the detailed-balance condition in Eq.~(\ref{eqn:detailed_balance}). Within this commuting sector the expression is consistent with the results of Ref.~\cite{summer2025resourcetheoreticalunificationmpemba}, where a similar expansion is used to highlight the role of the slowest eigenmode.

\subsection{Heating vs. cooling}

Heating and cooling protocols fundamentally differ in their initial system inverse temperature, $\beta_0$, relative to the final inverse temperature of the bath, $\beta$. 
To compare both processes, we consider the system evolving from a cold (hot) state, at $\beta_0 = \beta_\mathrm{C}$ $(\beta_0 = \beta_\mathrm{H})$, to a hot (cold) state at $\beta = \beta_\mathrm{H}<\beta_\mathrm{C}$ $(\beta = \beta_\mathrm{C})$ in the heating (cooling) protocol.
All quantities related to heating (cooling) are denoted with a superscript $\mathrm{H}$ $(\mathrm{C})$.

Regarding Eq.~(\ref{eqn:entropy_production_slowest}), three relevant factors must be considered when comparing thermalization processes. The first is the spectral gap, which, in the weak-coupling limit, is controlled by the dissipative rates of the Lindbladian \cite{can2019,claeys2022,costa2023}. 
These decay rates are determined by the coefficients in Eq.~(\ref{eqn:lindblad_operators}).
Under assumption (A), Eq.~(\ref{eqn:assumption_gap}), we have
\begin{equation}\label{eqn:spectral_gap_comparison}
    \left|\text{Re}\left(\lambda_1^\mathrm{H}\right)\right| > \left|\text{Re}\left(\lambda_1^\mathrm{C}\right)\right|.
\end{equation}
Consequently, the thermalization timescale is inherently shorter for heating than for cooling
\begin{equation}\label{eqn:timescale_comparison}
    \tilde{t}_\mathrm{H} = \frac{1}{\left|\text{Re}\left(\lambda_1^\mathrm{H}\right)\right|} < \tilde{t}_\mathrm{C} = \frac{1}{\left|\text{Re}\left(\lambda_1^\mathrm{C}\right)\right|}.
\end{equation}
For the models studied in Ref.~\cite{tejero2025}, numerical evidence is consistent with (A).

The spectral gap alone is not sufficient to determine the speed of convergence. The system may approach equilibrium either by cooling down or by heating up to reach the stationary state; nonetheless, the asymmetry between the two processes persists, even though the underlying dynamics are identical \cite{lapolla2020,ibanez2024}.
The second factor is the overlap of the initial state with the slowest decaying mode. 
Since $\Lambda_1^{\ell}$ is orthogonal to the stationary state, $\tr(\Lambda_1^{\ell}\tau_\beta)=0$, and the overlap can be written as $c_1=\tr[\Lambda_1^{\ell}(\tau_{\beta_0}-\tau_\beta)]$.
That is, it measures the projection of the difference between the initial and final thermal states onto the slowest left eigenmatrix, which in turn depends on the bath temperature through the Lindbladian. 
Unlike the spectral gap, whose ordering is assumed in (A), we find no model-independent ordering between $|c_1^\mathrm{H}|$ and $|c_1^\mathrm{C}|$ from this structure alone.
For the thermal qubit of Sec.~\ref{sec:thermal_qubit}, the overlap is known in closed form, see Eq.~(\ref{eqn:c1}). 
It represents the difference between the ground-state populations of the initial and final thermal states. 
Therefore, it is antisymmetric under the exchange of the initial and final temperatures. 
Hence $c_1^\mathrm{C}=-c_1^\mathrm{H}$, and $|c_1^\mathrm{H}|=|c_1^\mathrm{C}|$ exactly for any $\beta_\mathrm{H}$ and $\beta_\mathrm{C}$, so that in this model the overlap does not contribute to the asymmetry. 
Whether an overlap asymmetry arises in other systems has to be assessed case by case, and we do not rely on it in what follows.

The third factor in Eq.~(\ref{eqn:entropy_production_slowest}) is the weight $W_1$ of the slowest mode. 
It involves the inverse of the final thermal state, and therefore also depends on the bath temperature. 
For small temperature differences, the eigenmatrices $\Lambda_1^{r,\mathrm{H}}$ and $\Lambda_1^{r,\mathrm{C}}$ correspond to similar relaxation modes, so that $W_1^\mathrm{H}\approx W_1^\mathrm{C}$ to leading order in the quench. 
However, the heating--cooling asymmetry is itself a correction beyond that leading order, just because it vanishes in the linear-response regime. 
This implies that the difference between the two weights is of the same order as the asymmetry and cannot be ruled out beforehand.
Note that, for the thermal qubit, $W_1=2\cosh^2(\beta\hbar\omega/2)$, which is larger for the colder final state.

Within the quadratic approximation, the ratio of the initial entropy production rates of heating and cooling, $\sigma_0\equiv\sigma(0)$, is therefore
\begin{equation}\label{eqn:ratio_modes}
    \frac{\sigma_{0}^{\mathrm{H}}}{\sigma_{0}^{\mathrm{C}}}
    = \frac{\left|\text{Re}\left(\lambda_1^\mathrm{H}\right)\right|}{\left|\text{Re}\left(\lambda_1^\mathrm{C}\right)\right|}
      \left(\frac{|c_1^\mathrm{H}|}{|c_1^\mathrm{C}|}\right)^{2}
      \frac{W_1^\mathrm{H}}{W_1^\mathrm{C}} .
\end{equation}
Only the first factor is larger than one, and only under assumption (A), see Eq.~(\ref{eqn:spectral_gap_comparison}).
The overlap and weight factors depend on the system, and no general ordering of $\sigma_0^\mathrm{H}$ and $\sigma_0^\mathrm{C}$ follows from the expansion alone. 
In particular, the quadratic truncation is controlled by the size of the quench, whereas the asymmetry is a higher-order effect in the quench, so that the terms of third order in $\tau_\beta^{-1}\Delta\rho$ neglected in Eq.~(\ref{eqn:relative_eigenmatrices}) contribute at the same order as the asymmetry itself. 
For the thermal qubit, for instance, Eq.~(\ref{eqn:ratio_modes}) with the weights given above differs from the exact ratio, Eq.~(\ref{eqn:qho_quotient}), at first order in the temperature difference. 
The meaning of the expansion is thus to identify the ingredients that may produce an asymmetry, rather than to rank the two processes quantitatively. 
It is worth noticing that the exact solution of the thermal qubit presented in Sec.~\ref{sec:thermal_qubit}, which is free of any restriction on the quench size, shows that $\sigma_0^\mathrm{H}>\sigma_0^\mathrm{C}$ for arbitrary $\beta_\mathrm{H}<\beta_\mathrm{C}$ and that, in this model, the asymmetry is carried entirely by the spectral gap. 
This is consistent with findings for classical systems \cite{vanvu2021,ibanez2024} and with numerical results for quantum systems \cite{tejero2025}.
For $t>0$, the exponential decay $\exp[-2 |\text{Re}(\lambda_1)| t]$ in Eq.~(\ref{eqn:entropy_production_slowest}) is faster for heating due to the larger spectral gap, leading to more rapid convergence to the stationary state $\tau_{\beta_\mathrm{H}}$ compared with the cooling process toward $\tau_{\beta_\mathrm{C}}$.

\section{Thermal qubit}\label{sec:thermal_qubit}
We now apply the calculations and concepts discussed above to a qubit weakly coupled to a thermal bath, allowing it to thermalize at a different temperature. This simple model enables explicit computation of the entropy production rate, verification of the asymmetry between heating and cooling, and illustration of the implications of the uncertainty relations.

\subsection{Dynamics and entropy production}
Consider a two-level system described by the Hamiltonian $H = \hbar \omega\, \ket{1}\bra{1} = \hbar\omega(\mathrm{Id}-\sigma_z)/2$, where $ \sigma_z = \ket{0}\bra{0} - \ket{1}\bra{1}$. The energy of the ground state, $ \ket{0}$, is set to $0$ for simplicity, so $ H \ket{1} = \hbar \omega \ket{1}$.
The thermal state is directly given by
\begin{equation}\label{eqn:qho_gibbs}
    \tau_\beta = \dfrac{e^{-\beta H}}{Z} =  \frac{1}{Z} \left( \ket{0}\bra{0} + e^{-\beta \hbar \omega} \ket{1}\bra{1}\right),
\end{equation}
where $Z = 1 + e^{-\beta\hbar\omega}$ is the partition function. The jump operators are defined as in Eq. (\ref{eqn:lindblad_operators}), with $A=\sigma_-=\ket{0}\bra{1}$ and $A^\dagger=\sigma_+ =\ket{1}\bra{0}$. 
To simplify the notation, we consider the Fock-Liouville representation for the state of the system. Then, the thermal state of the system is given by a vector of the diagonal elements of the density matrix, $\rho=(\rho_{00} \ \rho_{11})^T$, corresponding to the occupation probabilities of the two states, with $\rho_{00}+\rho_{11}=1$. The off-diagonal elements have been removed as they are equal to zero. Since the Lindbladian is straightforwardly
\begin{equation}
    \mathcal{L} = \begin{pmatrix} -\gamma n_{\mathrm{B}}(\beta) & \gamma [n_{\mathrm{B}}(\beta) + 1] \\ \gamma n_{\mathrm{B}}(\beta) & -\gamma [n_{\mathrm{B}}(\beta) + 1] \end{pmatrix},
\end{equation}
where $\beta$ is the temperature of the thermal bath, the calculation of its eigenvalues and eigenvectors is analytically tractable. This allows us to write the evolution of the state evolution from $\beta_0$ to $\beta$ as
\begin{equation}
    \rho(t) = \tau_{\beta} + c_1 e^{- \gamma [1 + 2 n_{\mathrm{B}}(\beta)] t} \Lambda_1^r,
\end{equation}
where 
\begin{equation}\label{eqn:c1}
      c_1 = \frac{e^{\beta_0 \hbar \omega} - e^{\beta \hbar \omega}}{(1 + e^{\beta \hbar \omega})(1 + e^{\beta_0 \hbar \omega})}
\end{equation}
is the overlap with the initial state and $\Lambda_1^r = \left( 1 \ -1\right)^T$ is the right eigenvector corresponding to the decay mode. 
Equation~(\ref{eqn:c1}) can be written as $c_1=p_0(\beta_0)-p_0(\beta)$, with $p_0(\beta)=1/(1+e^{-\beta\hbar\omega})$ the ground-state population of $\tau_\beta$. Note that its numerator changes sign upon exchange $\beta_0\leftrightarrow\beta$, while its denominator is symmetric. 
Therefore, for a heating quench $(\beta_0=\beta_\mathrm{C},\beta=\beta_\mathrm{H})$ and the corresponding cooling quench ($\beta_0=\beta_\mathrm{H}$, $\beta=\beta_\mathrm{C}$) one has $c_1^\mathrm{C}= - c_1^\mathrm{H}$, i.e., $|c_1^\mathrm{H}|/|c_1^\mathrm{C}|=1$, for arbitrary $\beta_\mathrm{H}$ and $\beta_\mathrm{C}$. 
The overlap with the slowest decay mode is thus identical in modulus for both processes and does not contribute to the asymmetry in the thermal qubit. 
The decay rate in the exponent, $\Gamma(\beta)\equiv\gamma[1+2n_\mathrm{B}(\beta)]=|\text{Re}(\lambda_1)|$, i.e. the spectral gap, is the only element that differs between the two processes.
The entropy production rate is directly
\begin{equation}
\begin{split}
    \sigma(t) &= -\kb\frac{d}{dt} D(\rho(t) || \tau_{\beta}) \\
    &= -\kb\tr\left\{ \mathcal{L}[\rho(t)] (\ln \rho(t) - \ln \tau_{\beta}) \right\},
\end{split}
\end{equation}
where we have used the invariance of the trace, $\tr[\dot{\rho}(t)] = 0$. 

We now restrict the calculation to the entropy production rate at the initial time, $t=0$. 
For an initial thermal state $ \tau_{\beta_0}$, and a target state $\tau_{\beta}$, the entropy production rate reads
\begin{equation}\label{eqn:sigma_0}
    \sigma_0 = -\kb\tr\left\{ \mathcal{L}[\tau_{\beta_0}] ( \ln \tau_{\beta_0} - \ln \tau_{\beta} ) \right\}.
\end{equation}
Noting that the following term vanishes,
\begin{equation}
\tr\left\{\mathcal{L}[\tau_{\beta_0}] \left(\ln Z_{\beta_0} - \ln Z_{\beta}\right)\right\} = 0,
\end{equation}
and using $\ln\tau_{\beta_0}-\ln\tau_{\beta}=-(\beta_0-\beta)H+\text{constant}$, Eq.~(\ref{eqn:sigma_0}) reduces to 
\begin{equation}
    \sigma_0=\kb(\beta_0-\beta)\tr\{H\mathcal{L}[\tau_{\beta_0}]\},
\end{equation}
i.e., the initial heat flow weighted by the difference of the inverse temperatures. For the qubit we have, $\mathcal{L}[\tau_{\beta_0}]=-\Gamma(\beta)c_1 (1\ \ {-1})^T$, so that
\begin{equation}\label{eqn:sigma0_gamma_c1}
    \sigma_0 = \kb\hbar\omega(\beta_0-\beta)\Gamma(\beta)c_1 ,
\end{equation}
with $\Gamma(\beta)=\gamma[1+2n_\mathrm{B}(\beta)]$. Using Eq. (\ref{eqn:c1}), this can be written in closed form, depending only on the initial and final temperatures, as
\begin{equation}\label{eqn:dot_sigma_qho}
    \sigma_0 = \kb\gamma  \hbar \omega  \Delta \beta\,\frac{n_\mathrm{B}(\beta_0) - n_\mathrm{B}(\beta)}{1 + 2 n_\mathrm{B}(\beta_0)},
\end{equation}
where $\Delta \beta = \beta - \beta_0$.
This equation, particularized for the heating and cooling cases, reads
\begin{equation}
\begin{split}
    \sigma_{0}^{\mathrm{H}} & = \kb\gamma \hbar \omega  \Delta \beta_{\mathrm{C}\to\mathrm{H}}  \,\frac{ n_\mathrm{B}(\beta_\mathrm{H}) - n_\mathrm{B}(\beta_\mathrm{C}) }{1 + 2 n_\mathrm{B}(\beta_\mathrm{C})},\\
    \sigma_{0}^{\mathrm{C}} & = \kb\gamma \hbar \omega \Delta \beta_{\mathrm{C}\to\mathrm{H}}   \,\frac{n_\mathrm{B}(\beta_\mathrm{H}) - n_\mathrm{B}(\beta_\mathrm{C})}{1 + 2 n_\mathrm{B}(\beta_\mathrm{H})}.
\end{split}
\end{equation}
where $\Delta \beta_{\mathrm{C}\to\mathrm{H}} \equiv \beta_\mathrm{C} - \beta_\mathrm{H}$.
Comparing both entropy production rates, the only difference lies in the factor in the denominator, obtaining
\begin{equation}\label{eqn:qho_quotient}
    \dfrac{\sigma_{0}^{\mathrm{H}}}{ \sigma_{0}^{\mathrm{C}}} = \dfrac{1 + 2 n_\mathrm{B}(\beta_\mathrm{H})}{1 + 2 n_\mathrm{B}(\beta_\mathrm{C})}.
\end{equation}
Since $1 + 2 n_\mathrm{B}(\beta_\mathrm{H}) > 1 + 2 n_\mathrm{B}(\beta_\mathrm{C})$, it follows that $\sigma_{0}^{\mathrm{H}} > \sigma_{0}^{\mathrm{C}}$. 
This fact confirms faster entropy production during heating, i.e., faster thermalization toward a hot state than toward a cold one.
As seen in Eq.~(\ref{eqn:sigma0_gamma_c1}), the product $(\beta_0-\beta)c_1$ is invariant under the exchange of heating and cooling, because both factors change sign. The ratio of initial entropy production rates reduces to the spectral-gap ratio $\Gamma(\beta_\mathrm{H})/\Gamma(\beta_\mathrm{C})$, Eq.~(\ref{eqn:qho_quotient}). 
In the thermal qubit, we find that the asymmetry arises exclusively from the bath-temperature dependence of the dissipative rate.

By tracing the origin of the asymmetry, the present analysis provides a fundamental physical justification for the results in \cite{tejero2025}. In the particular case near thermodynamic equilibrium, where the system response is linear in the small temperature difference, heating and cooling dynamics exhibit no asymmetry, as expected. This is evident from Eq.~(\ref{eqn:qho_quotient}): for $n_\mathrm{B}(\beta_\mathrm{H}) \approx n_\mathrm{B}(\beta_\mathrm{C})$, the entropy production rates for heating and cooling become equal.

\subsection{Uncertainty relations}\label{subsec:uncertainty}
While the entropy production rate derived above explains the \emph{speed} asymmetry, thermodynamic uncertainty relations reveal its implications for the \emph{precision and stability} of the heat current. We now show how the TKUR is the natural complement to the entropy production analysis, completing the physical picture of asymmetric thermalization.
The concrete case of a thermal qubit allows us to explore the thermodynamic uncertainty relation and its role in determining the precision of heating and cooling processes. Since the system is coupled to a bath during thermalization, we focus on the bounds for the relative fluctuations of the heat current.

Focusing first on the KUR, the relevant quantity is the dynamical activity, which for quantum systems is given by \cite{vanvu2025}
\begin{equation}\label{eqn:activity}
\alpha = \sum_k \tr\left(L_k \tau_{\beta} L_k^\dagger\right),
\end{equation}
where $\tau_{\beta}$ is the final stationary state, and $L_k$ are the Lindblad operators. In our case of interest, Eq. (\ref{eqn:activity}) simplifies to a sum containing only two terms
\begin{equation}\label{eqn:activity_calc}
\begin{split}
\alpha &= \tr\left(L_1 \tau_{\beta} L_1^\dagger\right) + \tr\left(L_2 \tau_{\beta} L_2^\dagger\right)\\
& = \gamma \dfrac{2 n_{\mathrm{B}}(\beta) [1 + n_{\mathrm{B}}(\beta)]}{1 + 2 n_{\mathrm{B}}(\beta)},
\end{split}
\end{equation}
where $L_1$ and $L_2$ are given by Eq. (\ref{eqn:lindblad_operators}) for $A=\sigma_-$.
Note that this quantity depends solely on the temperature of the final state, through $n_{\mathrm{B}}$. The activity is a monotonically increasing function of $n_{\mathrm{B}}$ and since $n_{\mathrm{B}}>0$ for all $\beta$, it remains strictly positive.
Comparing the heating and cooling cases, we find that the activity is inherently larger for the heating case $\alpha_\mathrm{H} > \alpha_\mathrm{C}$.
\begin{figure}[t]
    \centering
    \includegraphics[width=1\linewidth]{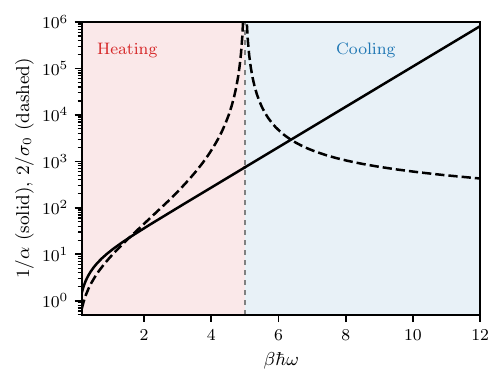}
    \caption{Inverse dynamical activity $1/\alpha$ (solid) and inverse initial entropy production rate $2/\sigma_0$ (dashed), Eq.~(\ref{eqn:dot_sigma_qho}), for the thermalization of the thermal qubit. The red region corresponds to heating and the blue region to cooling. The initial inverse temperature is fixed at $\beta_0=5$ (vertical dashed line); arrows indicate the direction of the quench (heating to the left, cooling to the right). Units are chosen such that $\hbar\omega=1$ and $\gamma=0.1$; consequently $\beta$ is measured in units of $1/(\hbar\omega)$ and the rates are measured in units of $\gamma$. The apparent divergence of $2/\sigma_0$ at $\beta=\beta_0$ reflects the vanishing of the entropy production (and of the asymmetry) in the linear-response regime.}
    \label{fig:activity_sigma}
\end{figure}
With both the entropy production rate and the dynamical activity at hand, we can use the TKUR, Eq.~(\ref{eqn:tkur}), to compare these quantities and determine which, between the inverse entropy production rate or the inverse activity, provides the tighter bound.
The comparison depends on the choice of parameters and the specific regimes considered in Eqs.~(\ref{eqn:dot_sigma_qho}) and (\ref{eqn:activity_calc}). Figure~\ref{fig:activity_sigma} shows both quantities for the thermal qubit, distinguishing between heating and cooling cases for fixed $\beta_0, \hbar\omega$, and $\gamma$. 
In general, the inverse entropy production rate is larger than the inverse activity at intermediate temperatures, whereas $1/\alpha$ is larger than $2/\sigma_0$ in the low-temperature and high-temperature limits.
For near temperature protocols, where $\beta_0$ and $\beta$ are similar, the inverse entropy is larger than the inverse activity. 

\subsection{Implications for asymmetry}\label{subsec:implications}
A few remarks are in order concerning the application of the TKUR/KUR to the present transient case. 
The quantum TKUR of Ref.~\cite{vanvu2025} is derived for time-integrated currents under GKSL dynamics, with the entropy production and dynamical activity evaluated in the stationary state. 
Uncertainty relations valid for finite observation times, arbitrary initial states or time-dependent dynamics have been derived both for classical systems \cite{liu2020tur,koyuk2020tur} and for Markovian open quantum systems \cite{vanvu2022precision}.
A transient version of the response bound discussed below is also given in Ref.~\cite{vanvu2025}. 
These formulations constrain the fluctuations of a current accumulated over a finite time interval $[0,\tau]$ in terms of quantities integrated along a trajectory, such as the entropy production $\Sigma_\tau=\int_0^\tau\sigma(t)dt$ or the dynamical activity $\mathcal{A}_\tau=\int_0^\tau\sum_k\tr[L_k\rho(t)L_k^\dagger]dt$, rather than in terms of instantaneous or stationary rates. Note that, for the relaxation process considered here, the entropy production accumulated over the whole process is just $\kb D(\tau_{\beta_0}||\tau_{\beta})$.

In the present work we evaluate the initial entropy production rate $\sigma_0$ on the initial thermal state and the dynamical activity $\alpha$ on the target thermal state, while the heat current itself is associated with the transient process. 
This mixed evaluation provides a convenient diagnostic of the relative strength of the two bounds. However, it does not constitute a rigorous application of a transient TUR/KUR.
A full formulation in the transient regime would require integrating the relevant rates along the entire trajectory. Such an analysis lies beyond the scope of the present article. The comparison performed here should therefore be understood as a qualitative indication of how the thermodynamic and kinetic constraints differ between heating and cooling.

Bearing this in mind, both the activity and the (initial) entropy production are larger for heating than for cooling. 
Consequently the lower bounds $2/\sigma$ and $1/\alpha$ that appear in the TKUR are lower for the heating process. That is, the bound on the relative fluctuations is less restrictive and leaves room for a higher precision. 
A lower bound does not by itself guarantee that the actual relative fluctuation $F_\phi$ is smaller. Establishing the latter would require an explicit evaluation of the heat current variance during the transient regime. The present results are therefore compatible with a higher precision during heating, yet they do not establish it.

Another bound of interest concerns the response of observables, such as the heat current $\phi$, to perturbations in the parameters of the Lindblad operators \cite{vanvu2025}:
\begin{equation}
\frac{\| \nabla \langle f(\phi) \rangle \|_1^2}{\text{Var}[f(\phi)]} \leq \tau \alpha.
\end{equation}
It quantifies how sensitive the observable $f(\phi)$ is to changes in parameters such as the coupling strength $\gamma$, normalized by its fluctuations. The larger $\alpha_{\mathrm{H}}$ in heating allows a stronger response, indicating that heating dynamics are more sensitive to changes in the system-bath interaction. 
This heightened responsiveness reflects the stronger dissipative coupling that accompanies the higher bath temperature. In the case of the transient problem considered here, the rigorous version of this bound replaces the right-hand side product $\tau\alpha$ by the activity integrated along the actual trajectory, $\mathcal{A}_\tau$ \cite{vanvu2025}, so that the stationary $\alpha$ used here should again be read as a diagnostic.

Regarding the quantum correction $\delta_\phi$ that appears in Eq.~(\ref{eqn:tkur}), $\delta_\phi=0$ recovers the classical limit \cite{vanvu2025}, whereas for $-2<\delta_\phi<0$ the factor $(1+\delta_\phi)^2$ is smaller than one.
In such case, the classical bound can be violated, i.e., quantum coherent dynamics can enhance the precision of the current beyond the classical bound.
In the present case both the initial and final states, as well as the evolved state under the Markovian dynamics, remain diagonal in the energy eigenbasis. This means that no off-diagonal coherences are generated and $\delta_\phi$ vanishes. 
The asymmetry observed here is thus driven purely by the dissipative rates. In systems with non-diagonal initial states or coherent driving, a non-zero $\delta_\phi$ could modify the bound and thereby the comparison between heating and cooling. An analysis of that regime, however, requires the extension of the perturbative treatment discussed to accommodate the non-commuting sector in Sec.~\ref{subsec:approximated_entropyprod}.

These considerations highlight a thermodynamic cost associated with the asymmetry: the larger entropy production in heating reflects greater irreversible dissipation. 
A possible trade-off between speed and efficiency, mediated by the uncertainty relations, may offer a guide for optimizing quantum thermodynamic devices in which the asymmetry may be engineered to accelerate heating cycles at the expense of thermodynamic efficiency.

\section{Conclusion}
In this paper, we have analyzed the physical mechanism underlying the asymmetry between heating and cooling in open quantum systems that thermalize under Davies dynamics while remaining diagonal in the energy basis. 
By means of a spectral decomposition of the Liouvillian together with the exact solution of the thermal qubit and the quantum TKUR, we restrict our evaluation of the entropy production rate to the commuting sector. 
Through this analysis, we establish why heating induces faster thermalization than cooling and identify the precise physical quantities governing this asymmetry.

Restricting to such commuting sector, the expansion isolates three ingredients of the entropy production rate: the spectral gap, the overlap of the initial state with the slowest relaxation mode, and a mode weight. 
Only the spectral gap is ordered between heating and cooling, under assumption (A), and up to a quadratic term. 
Particularly for the thermal qubit, the ratio of the heating to cooling rates admits an exact closed form, which shows that the higher initial entropy production in heating is due entirely to the larger spectral gap, since the overlaps of both processes are equal in modulus. 
The TKUR is then employed as a diagnostic tool. The larger entropy production and dynamical activity associated with heating lower the bounds on relative heat-current fluctuations. A lower bound does not by itself prove that the actual fluctuations are smaller, but it is consistent with a more precise heating process and points to an interesting open question concerning the saturation of transient uncertainty relations.

These findings suggest a possible trade-off between speed and thermodynamic efficiency, of potential relevance for quantum thermal machines, which we have not quantified here.
The expansion in Sec.~\ref{subsec:approximated_entropyprod} is general, applying to any Markovian dynamics that thermalize under detailed balance and remain within the diagonal sector.
Nevertheless, the relative importance of the spectral gap, overlap terms, and weight factors must be evaluated for each specific system. 
This means that the present TKUR analysis should be understood as a qualitative diagnostic rather than a rigorous transient bound.
Future research may extend this framework to non-commuting regimes involving coherent driving, non-diagonal initial states, multiple thermal baths, or non-Markovian memory effects, where new mechanisms underlying the asymmetry may be revealed.

\section{Acknowledgements}
The author would like to thank Daniel Manzano, Michalis Skotiniotis, Joe Dunlop, and Massimiliano Sacchi for their insightful comments.
We acknowledge funding from the Ministry for Digital Transformation and of Civil Service of the Spanish Government through projects, PID2021-128970OA-I00, PID2023-149365NB-I00, PID2024-162155OB-I00, FPU20/02835 and QUANTUM ENIA project call - Quantum Spain project, and by the European Union through the Recovery, Transformation and Resilience Plan - NextGenerationEU within the framework of the Digital Spain 2026 Agenda, and also the FEDER/Junta de Andalucía program A.FQM.752.UGR20. Finally, we are also grateful for the technical support provided by PROTEUS, the supercomputing center of the Institute Carlos I for Theoretical and Computational Physics in Granada, Spain.

\bibliography{bibliography}
\end{document}